\documentclass[twocolumn,aps,preprintnumbers,amsmath,amssymb,superscriptaddress,footinbib]{revtex4}

\usepackage{graphicx,subfigure,multirow}
\usepackage{longtable}
\usepackage{dcolumn}
\usepackage{bm}
\usepackage[all]{xy}
\usepackage{color}
\usepackage[usenames,dvipsnames]{xcolor}
\usepackage[unicode=true,pdfusetitle,
 bookmarks=true,bookmarksnumbered=false,bookmarksopen=false,citecolor=Turquoise,
 breaklinks=false,pdfborder={0 0 1},backref=false,colorlinks=true,pdfpagemode=FullScreen]
 {hyperref}

\usepackage{ulem}

\makeatletter

\newcommand{\fmslash}[2][0mu]{%
  \mathchoice
    {\fmsl@sh\displaystyle{#1}{#2}}%
    {\fmsl@sh\textstyle{#1}{#2}}%
    {\fmsl@sh\scriptstyle{#1}{#2}}%
    {\fmsl@sh\scriptscriptstyle{#1}{#2}}}
\newcommand{\fmsl@sh}[3]{%
  \m@th\ooalign{$\hfil#1\mkern#2/\hfil$\crcr$#1#3$}}
\makeatother

\newcommand{\lsim}{{\;\raise0.3ex\hbox{$<$\kern-0.75em\raise-1.1ex\hbox{$\sim$}}\;}}
\newcommand{\gsim}{{\;\raise0.3ex\hbox{$>$\kern-0.75em\raise-1.1ex\hbox{$\sim$}}\;}}

\newcommand{\beq}{\begin{equation}}
\newcommand{\eeq}{\end{equation}}
\newcommand{\bea}{\begin{eqnarray}}
\newcommand{\eea}{\end{eqnarray}}
\mathchardef\minus="002D

\usepackage{color}

\begin{document}
\title{Dark Matter ``Collider'' from Inelastic Boosted Dark Matter}

\author{Doojin Kim}
\email{doojin.kim@cern.ch}
\affiliation{Theory Division, CERN, CH-1211 Geneva 23, Switzerland}
\affiliation{Department of Physics, University of Florida, Gainesville, FL 32611, USA}
\author{Jong-Chul Park}
\email{jcpark@cnu.ac.kr}
\affiliation{Department of Physics, Chungnam National University, Daejeon 34134, Republic of Korea}
\author{Seodong Shin}
\email{shinseod@indiana.edu}
\affiliation{Department of Physics \& IPAP, Yonsei University, Seoul 03722, Republic of Korea}
\affiliation{Physics Department, Indiana University, Bloomington, IN 47405, USA}

\preprint{
\begin{minipage}[b]{1\linewidth}
\begin{flushright}
CERN-TH-2016-258\\
 \end{flushright}
\end{minipage}
}

\begin{abstract}
We propose a {\it novel} dark matter (DM) detection strategy for the models with non-minimal dark sector.
The main ingredients in the underlying DM scenario are a boosted DM particle and a heavier dark sector state.
The relativistic DM impinged on target material scatters off {\it in}elastically to the heavier state which subsequently decays into DM along with lighter states including visible (Standard Model) particles.
The expected signal event, therefore, accompanies a visible signature by the secondary cascade process associated with a recoiling of the target particle, differing from the typical neutrino signal {\it not} involving the secondary signature.
We then discuss various kinematic features followed by DM detection prospects at large volume neutrino detectors with a model framework where a dark gauge boson is the mediator between the Standard Model particles and DM.
\end{abstract}


\maketitle

\paragraph*{{\bf Introduction.}}
A tremendous amount of effort has been made to detect non-gravitational signals of dark matter (DM) in direct detection, indirect detection, and collider experiments.
Most of them, however, have provided strong constraints on DM models rather than unambiguous discovery signatures.
This fact has recently motivated research programs to study non-conventional DM scenarios such as secluded DM~\cite{Huh:2007zw, Pospelov:2007mp, Kim:2008pp, Kim:2009ke, Chun:2010ve, Park:2013bza, Belanger:2013tla, Kim:2016csm} and non-minimal dark sector models including assisted freeze-out~\cite{Belanger:2011ww}, boosted DM (BDM)~\cite{Agashe:2014yua, Berger:2014sqa, Kong:2014mia, Alhazmi:2016qcs}, dark cascade scenarios~\cite{Kim:2015usa, Kim:2015gka}, multi-component DM~\cite{DDM1, DDM2, DDMDD, DDMAMS, DDMLHC1, DDMLHC2, Boddy:2016fds, Boddy:2016hbp}, and DM transporting mechanism for cosmic-ray excesses~\cite{Kim:2017qaw}.

Promising strategies to examine those possibilities include the observation of relativistic DM scattering with targets in terrestrial experiments to overcome its small interaction with Standard Model (SM) particles. Relativistic DM particles may arise, for example, by the annihilation of a heavier DM pair (into a lighter DM pair) at the {\it present} universe in multi-component DM scenarios or the decay of an energetic mediator created via (proton or electron) beam bombardment on fixed target material. Therefore, large volume neutrino detectors~\cite{Agashe:2014yua} and fixed target experiments with high intensity~\cite{Batell:2009di, deNiverville:2011it,Izaguirre:2014dua,Battaglieri:2014qoa,Battaglieri:2016ggd} can probe relevant signals.
We remark that the search scheme in both cases is based on simple number counting over the expected number of background events.
Indeed, this scheme has intrinsic limitation: in particular for neutrino detectors, it is hard to discern a relativistic scattering signal of DM from a neutrino scattering event, the main {\it ir}reducible background.

\begin{figure}[t]
\centering
\includegraphics[width=5.5cm]{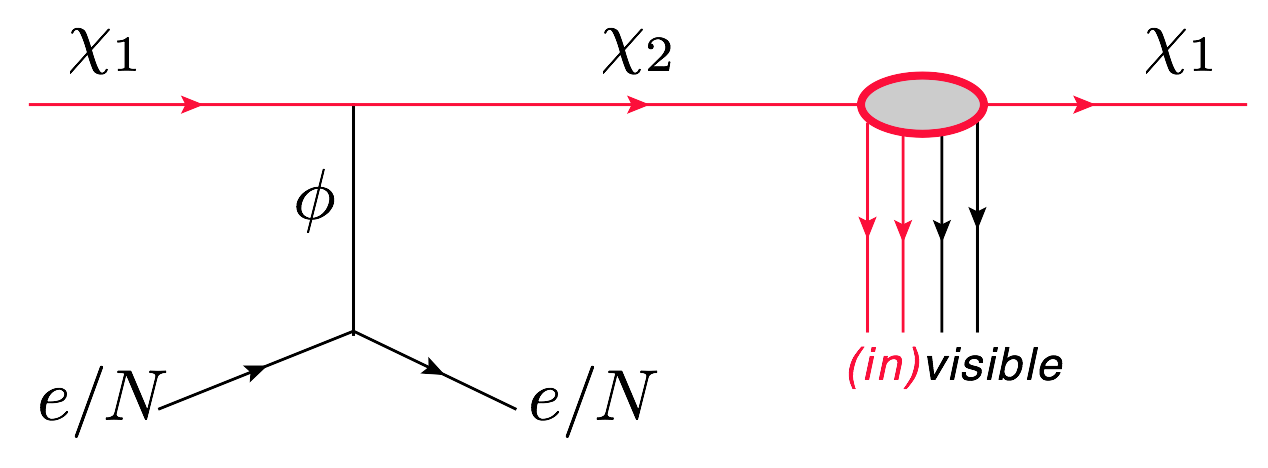}
\caption{\label{fig:scenario} Inelastic boosted DM direct detection scenarios under consideration.}
\end{figure}

In this letter, we propose a {\it novel} channel which takes a key role in search for the relativistic DM scattering signals arising in various models comprising an additional (unstable) dark sector particle.
This is schematically depicted in Fig.~\ref{fig:scenario} where the scattered dark sector particle (denoted by $\chi_2$) differs from the incoming DM (denoted by $\chi_1$), i.e., an inelastic scattering occurs in the recoil of the target.
Furthermore, $\chi_2$ is heavier than $\chi_1$ so that the former subsequently decays into lighter states including the latter and visible SM particles, which is reminiscent of typical cascade decay signatures in collider experiments.
The expected signal, therefore, involves a recoiling of target material and (visible) decay products from the secondary process of $\chi_2$~\cite{Izaguirre:2014dua}.
We {\it first} find that this feature is clearly distinctive from high energetic neutrino signatures in DM detection.\footnote{As a non-relativistic scattering, there is a scenario of inelastic excitation of DM followed by de-excitation via decay into photon, which can be tested in a neutrinoless double beta decay experiment~\cite{Pospelov:2013nea}.
The strategy may look similar to ours, but the process lacks the observation of the primary recoil signal by inelastic DM excitation and the monochromatic X-ray photon is the only secondary signal.} 

\paragraph*{{\bf Model framework.}}
To validate the DM scenario explained above, we first delineate a DM model framework which contains the cascade process depicted in Fig.~\ref{fig:scenario}.
Employing a Dirac fermionic DM $\chi_1$ for simplicity, we assume that it interacts with target SM particles (e.g., electron or nucleus) via a $t$-channel exchange of the mediator $\phi$.
As stated earlier, we further assume that the outgoing dark sector particle is {\it not} $\chi_1$ but a heavier unstable particle $\chi_2$ (i.e., $m_{\chi_2} > m_{\chi_1}$).
In principle, the mediator $\phi$ can be either a SM or a new physics particle, but we take a ``dark'' gauge boson $X_\mu$ for simplicity from the following toy model Lagrangian with a dark U(1)$_{\rm X}$ gauge symmetry:
\begin{align}
\mathcal{L}_X \supset -\frac{\sin\epsilon}{2} F_{\mu\nu}X^{\mu\nu}+g_{12} \bar{\chi}_2 \gamma^{\mu}\chi_1 X_{\mu} + h.c.\,, \label{eq:intX}
\end{align}
where the first term describes the kinetic mixing between U(1)$_{\rm X}$ and U(1)$_{\rm EM}$~\cite{Okun:1982xi, Galison:1983pa, Holdom:1985ag, Huh:2007zw, Chun:2010ve, Park:2012xq} parameterized by $\epsilon$.
The off-diagonal gauge interaction of $\chi_1$ and $\chi_2$ with $X_\mu$ appears in the second term with coupling $g_{12}$.
We expect that such a vertex may arise, e.g., from the mixing in the dark sector after imposing different U(1)$_{\rm X}$ charges to $\chi_1$ and $\chi_2$ (see also Ref.~\cite{Kim:2010gx} for the mixing in the SM quark sector with a U(1)$'$ symmetry).
More concrete model building including other possible scenarios (e.g., Higgs portal) will be available in our future work~\cite{future}.

The heavier nature of $\chi_2$ renders its decay eventually into $\chi_1$ plus SM or other (invisible) dark sector particles.
Such a decay, in general, proceeds via a sequential cascade process as symbolized by a red-circled blob in Fig.~\ref{fig:scenario}.
Hence, the expected signal event is featured by a recoil of the target SM particle, accompanying secondary signatures from the cascade decay process.
As a minimal choice, we take a single-step cascade decay of $\chi_2$ throughout this letter, i.e., $\chi_2$ decays back into $\chi_1$ and $\phi$ by the interactions in Eq.~\eqref{eq:intX}.

We first calculate the matrix element squared for the process $\chi_1 T \rightarrow \chi_2 T$ with $T$ being the associated target:
\begin{align}
\overline{\left|\mathcal{M} \right|}^2
&= \frac{8(\epsilon e g_{12})^2 m_T}{\{2m_T(E_{\chi_2}-E_{\chi_1})-m_{\phi}^2\}^2} \nonumber \\
&\hspace{0.2cm} \times \left[\mathcal{M}_0  (F_1 + \kappa F_2)^2   +  \mathcal{M}_1 \left\{ - (F_1 + \kappa F_2)\kappa F_2 \right. \right. \nonumber \\
&\hspace{0.4cm} \left. \left. \, +
(\kappa F_2)^2
(E_{\chi_1} - E_{\chi_2} + 2 m_T)/(4 m_T) \right\}\right]\,. \label{eq:matrixX}
\end{align}
Here $\mathcal{M}_0$ and $\mathcal{M}_1$
are defined as follows:
\begin{align}
\mathcal{M}_0 &= \left[m_T( E_{\chi_1}^2+E_{\chi_2}^2)-(\delta m_{\chi})^2(E_{\chi_2} -E_{\chi_1} +m_T)/2\right. \nonumber \\
&\hspace{0.3cm} +\left. m_T^2(E_{\chi_2}-E_{\chi_1})+m_{\chi_1}^2E_{\chi_2} - m_{\chi_2}^2 E_{\chi_1} \right]\,, \\
\mathcal{M}_1 &= m_T \left[ \left((E_{\chi_1} + E_{\chi_2}) -
(m_{\chi_2}^2 - m_{\chi_1}^2)/(2m_T) \right)^2 \right.  \\
&\left. +\, (E_{\chi_1} - E_{\chi_2} + 2 m_T) \left\{(E_{\chi_2} - E_{\chi_1} ) - (\delta m)^2/(2m_T) \right\} \right],  \nonumber
\end{align}
where $\delta m_{\chi} \equiv m_{\chi_2}-m_{\chi_1}$ and $E_{\chi_{1(2)}}$ is the $\chi_{1(2)}$ energy measured in the laboratory frame.\footnote{Obviously, for the fixed target experiments, the target frame is the same as the laboratory frame.}
For the two form factors $F_1$ and $F_2$, we set them to be 1 and 0 for the electron target (or $e$-scattering), whereas we employ nontrivial values as per Ref.~\cite{Qattan:2004ht} for the proton target (or $p$-scattering) together with the proton anomalous magnetic moment $\kappa=1.79$.

\paragraph*{{\bf Kinematic features.}}
We now discuss interesting kinematic features arising in the model framework discussed earlier.
Like ordinary colliders, the maximum mass reach of $\chi_2$ is $\sqrt{s}-m_T$ with $\sqrt{s}$ being the overall center-of-mass energy (i.e., $s=m_T^2+2E_{\chi_1}m_T+m_{\chi_1}^2$):
\begin{align}
m_{\chi_2}\leq \sqrt{m_T^2+2E_{\chi_1}m_T+m_{\chi_1}^2}-m_T\,.
\label{eq:mchi2}
\end{align}
If $\chi_1$ is much heavier than the target (i.e., $m_{\chi_1} \gg m_T$) along with a decent boost $\gamma_{\chi_1}$~\footnote{Note that $\gamma_{\chi_1}$ is the boost factor of $\chi_1$, i.e., $E_{\chi_1} = \gamma_{\chi_1} m_{\chi_1}$.}, the above relation is approximated to
\begin{align}
m_{\chi_2} \lesssim m_{\chi_1}+(\gamma_{\chi_1}-1)m_T\,,
\end{align}
to which our $e$-scattering corresponds.
On the other hand, the opposite limit, $m_{\chi_1} \ll m_T$, results in
\begin{align}
m_{\chi_2} \lesssim \gamma_{\chi_1}m_{\chi_1}\,,
\end{align}
allowing us to probe much heavier dark sector states than the incoming DM, which is possible for $p$-scattering.

\begin{figure}[t]
\centering
\includegraphics[width=0.485\linewidth]{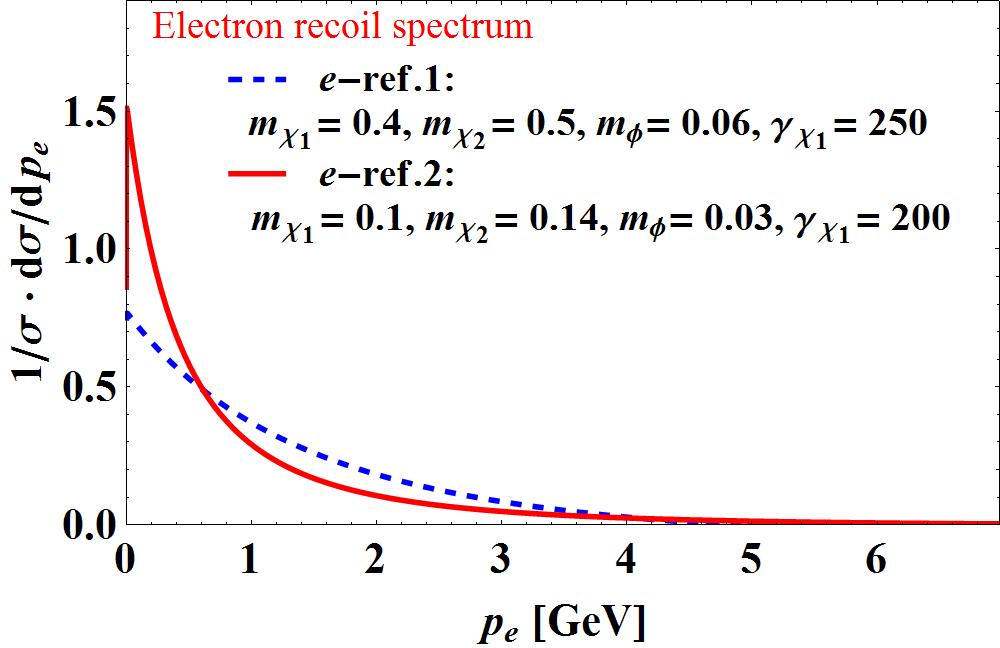}
\includegraphics[width=0.494\linewidth]{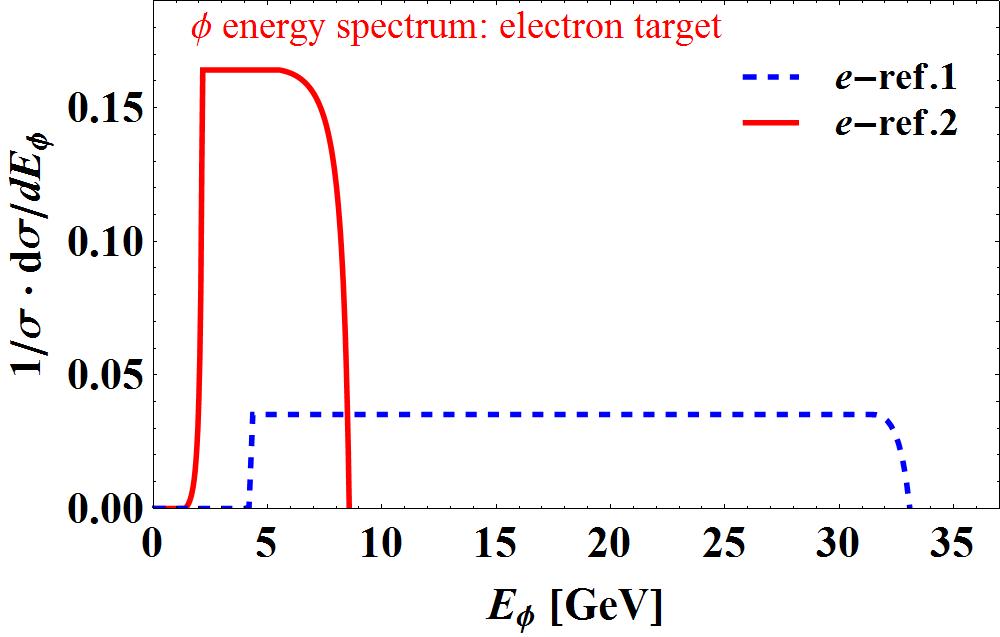}\\ \vspace{0.1cm}
\includegraphics[width=0.471\linewidth]{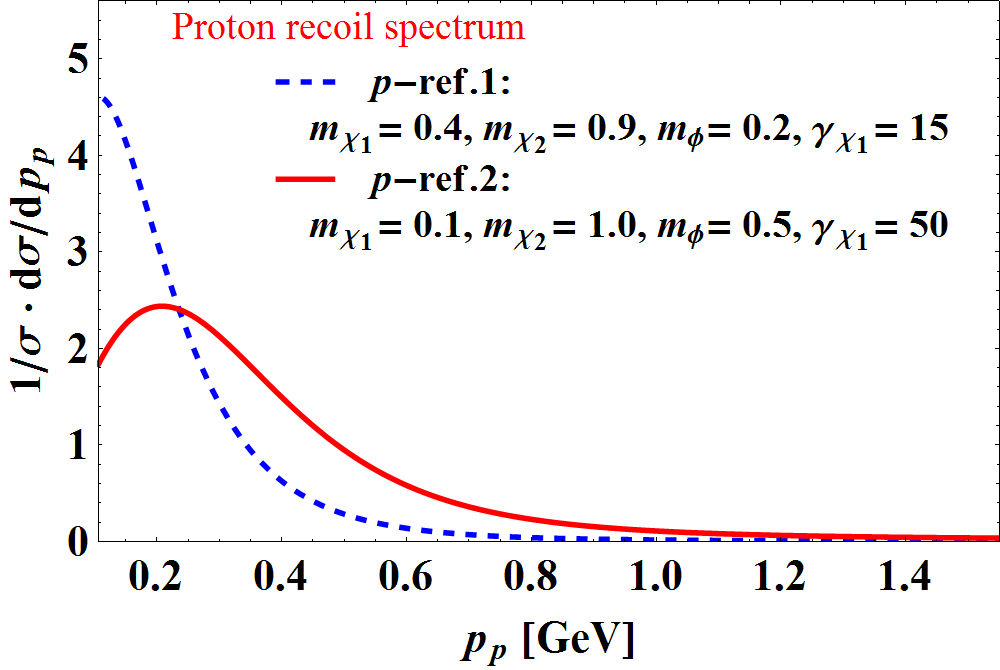}
\includegraphics[width=0.494\linewidth]{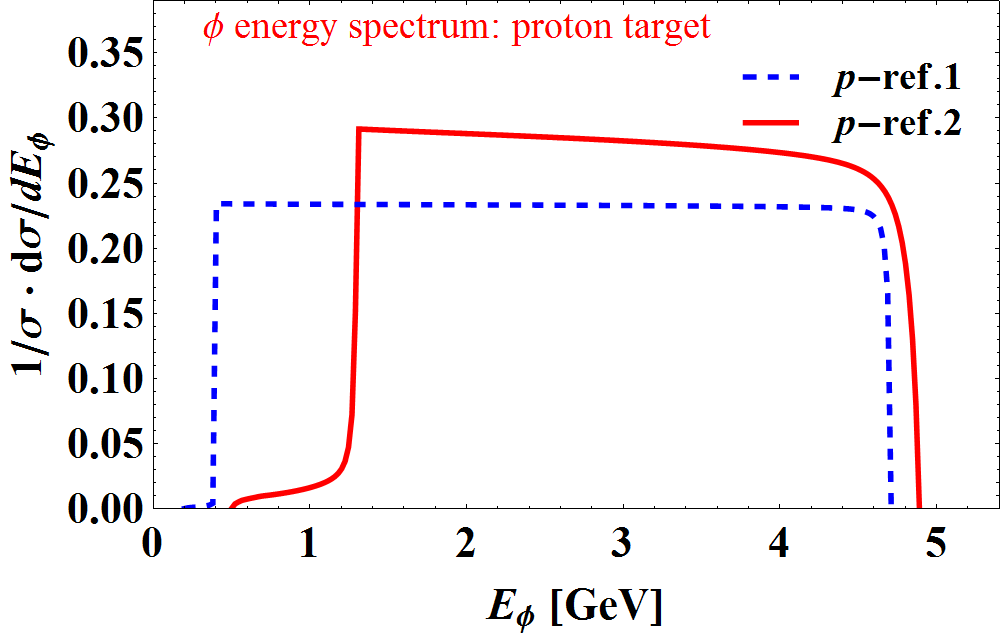}
\caption{\label{fig:spectrum} Expected unit-normalized energy spectra of the recoiling target particles from the primary vertex (left panels) and outgoing mediators from the secondary vertex (right panels) for $e$-scattering (top panels) and $p$-scattering (bottom panels).
The reference masses are in unit of GeV.
}
\end{figure}

We next discuss the expected energy spectra of the recoiling target and the visible particles from the secondary vertex.
In the laboratory frame, the differential cross section is
\begin{align}
\frac{d\sigma}{dE_T}=\frac{m_T }{8\pi\lambda(s,m_T^2,m_{\chi_1}^2)}\overline{\left|\mathcal{M} \right|}^2 \,, \label{eq:ETspec}
\end{align}
where $E_T$ is the energy of the recoiling target and $\lambda(x,y,z) = (x - y - z)^2 - 4yz$.
Here $\overline{\left|\mathcal{M} \right|}^2$ is expressed in terms of $E_{T} = E_{\chi_1}+m_T-E_{\chi_2}$.
Note that the recoil energy in usual direct detection experiments corresponds to the magnitude of the spatial momentum (equivalently, the kinetic energy) of the recoiling target, i.e., $p_T =  \sqrt{E_T^2 - m_T^2}$.
We find that kinematically allowed maximum (minimum) recoil energy $E_T^+$ ($E_T^-$) is 
\begin{align}
E_T^{\pm} = \frac{s+m_T^2-m_{\chi_2}^2}{2\sqrt{s}}\frac{E_{\chi_1}+m_T}{\sqrt{s}}\pm \frac{\lambda^{1/2}(s,m_T^2,m_{\chi_2}^2)}{2\sqrt{s}}\frac{p_{\chi_1}}{\sqrt{s}}\,,
\end{align}
where $p_{\chi_1} = \sqrt{E_{\chi_1}^2 - m_{\chi_1}^2}$\,.
The upper-left panel ($e$-scattering) and the lower-left panel ($p$-scattering) in Fig.~\ref{fig:spectrum} demonstrate expected unit-normalized recoil energy spectra for our four reference points ($e$-ref.1, $e$-ref.2, $p$-ref.1, and $p$-ref.2) as detailed in the plots, which are not only safe from various experimental bounds~\cite{Essig:2013lka, Banerjee:2016tad} but phenomenologically well-motivated~\cite{Huh:2007zw,Pospelov:2007mp,Batell:2009di, deNiverville:2011it,Izaguirre:2014dua,Battaglieri:2014qoa,Battaglieri:2016ggd,ArkaniHamed:2008qn,Chun:2008by}.
Note that the differential cross section is greater for the smaller momentum transfer as expected in Eq.~\eqref{eq:matrixX}.

The spectral behavior in the distribution of $\phi$ energy $E_\phi$, in principle, depends on the relevant vertex structure.
In our toy model, due to the vector-like nature of the mediator coupling, $\chi_2$ is produced in an unpolarized way, so that it can be treated {\it effectively} as a scalar.
For a simple two-body decay, the energy spectra of decay products have been extensively examined in the context of collider phenomenology~\cite{Agashe:2012bn, Agashe:2012fs, Agashe:2013eba, Chen:2014oha, Agashe:2015wwa, Agashe:2015ike, Agashe:2016bok} and cosmic-ray phenomenology~\cite{Kim:2015usa, Kim:2015gka, Boddy:2016fds, Boddy:2016hbp, 1971NASSP.249.....S}.
For generality, we consider the case that $m_\phi$ is not negligible, finding the following expression based on the formulation in Ref.~\cite{Agashe:2015ike}:
\bea
\frac{d\sigma}{dE_{\phi}}=\int d\gamma_{\chi_2}\frac{d\sigma}{d\gamma_{\chi_2}}\frac{1}{2E_{\phi}^*\sqrt{\gamma_{\chi_2}^2-1}}\,,
\eea
where $E_{\phi}^*$ is the $\phi$ energy measured in the $\chi_2$ rest frame.
The detailed expressions for the integral range are not illustrative, so we instead refer to Refs.~\cite{future,Agashe:2015ike}.
Here the boost distribution of $\chi_2$, $d\sigma/d\gamma_{\chi_2}$, can be easily obtained from the $\chi_2$ energy spectrum, which is, in turn, derived from Eq.~\eqref{eq:ETspec} with $E_T$ replaced by $E_{\chi_1}+m_T-E_{\chi_2}$.

The expected (unit-normalized) energy spectra of $\phi$ produced in the cascade process for our reference points are exhibited in the upper-right panel ($e$-scattering) and the lower-right panel ($p$-scattering) in Fig.~\ref{fig:spectrum}, respectively.
For the chosen reference points, we find that $E_{\chi_2}$ values are highly localized towards the kinematic endpoint, and therefore, the resulting $\phi$ energy spectrum appears almost box-like.
In practice, the energy of $\phi$ can be measured from its visible decay products. Throughout this letter, we assume that the mediator $\phi$ predominantly decays into $e^+e^-$. We then find the decay $\chi_2 \to \chi_1 \phi \to \chi_1 e^+ e^-$ occurs within at most $\lesssim$ 1\,cm for our reference points, which is below the detector resolutions identifying separate vertices (e.g., a few cm for DUNE~\cite{Acciarri:2016ooe}). Depending on $m_{\phi}$ and $\epsilon$, one may also consider an appreciable displaced vertex in the decay $\chi_2 \to \chi_1 e^+ e^-$ which can be observed at detectors with high vertex-position resolution (e.g., DUNE and SHiP)~\cite{future}. In either case, a signal event is characterized by a recoiling target ($e$ or $p$) and a $e^+e^-$ pair, so the angular separations among them would be critical to identify the signal events, which will be discussed in the next section.

\paragraph*{{\bf Detection prospects.}}

\begin{table}[t]
\centering
\begin{tabular}{c c c c c c}
Exp. & Volume [Mt] & $E_e^{\textnormal{th}}$ [GeV] & $E_p^{\textnormal{th}}$ [GeV] & ~~$\theta_e^{\textnormal{res}}$ [${}^{\circ}$]& ~~$\theta_p^{\textnormal{res}}$ [${}^{\circ}$] \\
\hline \hline
SK & 0.0224 & 0.1 & 1.07 & 3 & 3 \\
\hline
HK & 0.56 & 0.1 & 1.07 & 3 & 3 \\
DUNE & 0.04 & 0.03 & 0.05 & 1 & 5
\end{tabular}
\caption{\label{tab:exp} Summary of the volume, threshold energy, and
angular resolution of considered experiments from Refs.~\cite{Fechner:2009aa}, \cite{Kearns:2013lea}, and \cite{Acciarri:2015uup} for SK, HK, and DUNE, respectively.
$E_e^{\rm th}$ at SK/HK could be lowered below 0.1 GeV with worse angular resolution.
Angular resolution gets better with higher $p_T$.
}
\end{table}

Based on the signal features discussed so far, we are now in the position to assess the detection prospects of our signal.
In order for our signal to be sensitive even with small flux, we choose large volume neutrino detectors: Super-Kamiokande (SK), Hyper-Kamiokande (HK), and Deep Underground Neutrino Experiment (DUNE) where the latter two are future proposals.
We summarize their key attributes in Table~\ref{tab:exp}. While we do our analysis having in mind a BDM type scenario for obtaining boosted DM, we again emphasize that fixed target experiments (e.g., LBNF/DUNE~\cite{Acciarri:2016ooe}, SHiP~\cite{Anelli:2015pba}, and T2HKK~\cite{Abe:2016ero}) are alternative sources~\cite{future}. 

The energy and angular resolutions for $e$-scattering are usually better than those for $p$-scattering, especially in SK/HK.
This is because large momentum transfer above $m_p = 0.938$ GeV is required for the recoiling proton to produce Cherenkov radiation ($p_p \gtrsim 1.07$ GeV for SK/HK~\cite{Fechner:2009aa}).
Note that this requirement is rather relaxed in liquid Ar TPC detectors like DUNE.
Reminding the trend that the differential cross section is larger for smaller momentum transfer, one can expect that $e$-scattering is preferred over $p$-scattering in SK/HK if focusing only on the recoil signal.
For $p$-scattering, we further restrict ourselves to $p_p \lesssim 1.8$ GeV to avoid the possibility of deep inelastic scattering.

\begin{figure}
\centering
\includegraphics[width=0.49\linewidth]{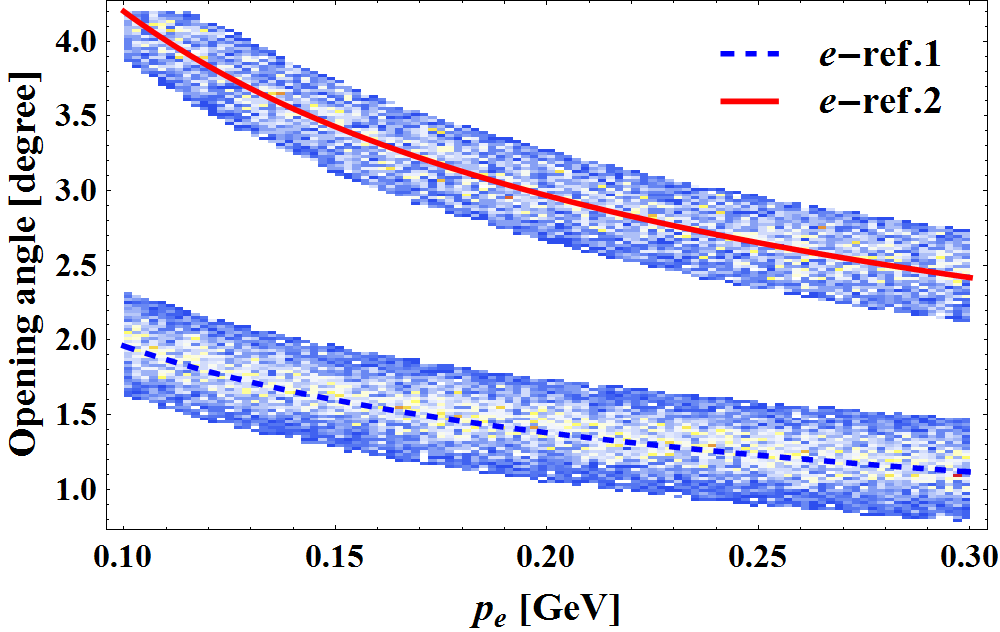}
\includegraphics[width=0.49\linewidth]{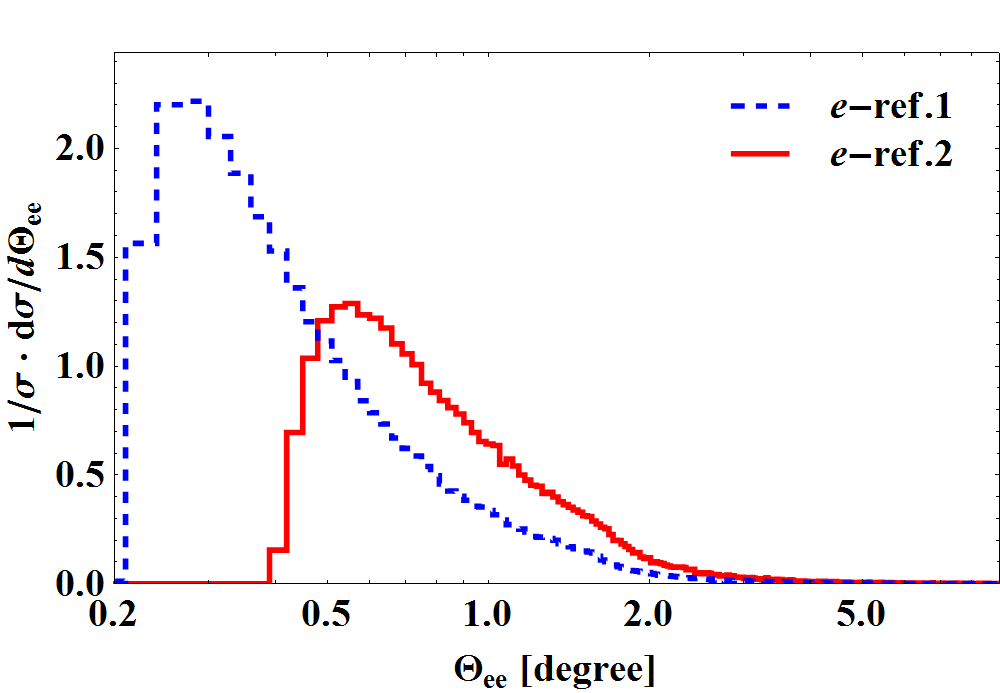}
\caption{\label{fig:angle}
Left panel: angular separation between the recoiling target and $\chi_2$ (red solid and blue dashed lines) or the mediator $\phi$ (temperature-scaled regions). Right panel: angular separation between the $e^+e^-$ pair from the $\phi$ decay.
}
\end{figure}

As stated before, observation of the secondary cascade signal plays the key role in discovery of our DM signal.
We point out that the visible particles are often collimated due to the large boost of the incident DM.
Therefore, unambiguous signal identification depends on what extent we can separate those (highly) collimated signals beyond the angular resolutions of the detectors.
Defining $\theta_{\chi_2}$ as the angle between the recoiling target and $\chi_2$ in the laboratory frame, we obtain
\begin{align}
\cos\theta_{\chi_2} = \frac{E_T E_{\chi_2} + (m_T^2 + m_{\chi_2}^2 - s) /2 }{\sqrt{(E_{\chi_2}^2 - m_{\chi_2}^2)(E_T^2 - m_T^2)} }\,.
\end{align}
The value $\theta_{\chi_2}$ roughly determines the angular separation between the primary and secondary signals when $\chi_2$, $\phi$, and the decay products ($e^+ e^-$) are highly collimated.
This is true for $e$-scattering as shown in the left panel of Fig.~\ref{fig:angle}.
The red solid and blue dashed lines (temperature-scaled bands) show the angle between the recoiling target and $\chi_2$ ($\phi$), from which we clearly see that $\chi_2$ and $\phi$ are collimated.
We further check that the angular separation between $e^+$ and $e^-$ from the $\phi$ decay are mostly within $1.5^{\circ}$ as shown in the right panel of Fig.~\ref{fig:angle}.
Adopting the angular resolution $3^{\circ}$ for SK/HK, we find that our reference point $e$-ref.2 manifests two separable signatures for most momentum values of the recoiling electron $p_e \in [0.1, 0.3]$ GeV (see also Table~\ref{tab:exp}).\footnote{For $e$-ref.1, detecting events with separable signatures is very challenging according to the left panel of Fig.~\ref{fig:angle}, which enforces us to perform a more careful analysis in regard to angular separation by considering both the recoil electron and the $e^+e^-$ pair from the $\phi$ decay. }

On the other hand, larger angular separation is possible for our $p$-scattering reference points because $E_{\chi_1}$, $m_p$, and $m_{\chi_2}$ are roughly of the same order so that typical $\chi_2$'s are neither too boosted nor too aligned along the recoiling proton direction.
In addition, we observe that the opening angle of $\phi$ decay products (defined as $\Theta_{ee}$) gets larger. Given a mediator boost factor $\gamma_{\phi}$, we find
\begin{align}
\Theta_{ee} \simeq \arccos
\left[1 - 2/(\gamma_{\phi}^2\sin^2\theta+\cos^2\theta) \right]\,,
\end{align}
where $\theta$ is the emission angle of one of the decay products with respect to the $\phi$ boost direction in the $\phi$ rest frame.
Here we took the fact that $m_{\phi} \gg m_e$ for all our reference points.
It is easy to see that the opening angle is greater than $6^{\circ}$ for all possible $\theta$ as far as $\gamma_{\phi} \lesssim 20$.
We then find that our reference points selected for $p$-scattering are anticipated to have three resolvable signatures in most of the allowed phase space, whereas those for $e$-scattering would involve two signatures.
This is an unarguable advantage of $p$-scattering although the cross section is smaller than that for $e$-scattering.

\begin{table}
\centering
\begin{tabular}{c c c c c c}
Exp. & Run time & ~$e$-ref.1 & ~$e$-ref.2 & ~~$p$-ref.1 & ~~$p$-ref.2\\
\hline \hline
SK & 13.6 yr & ~170 & ~7.1 & ~~3500 & ~~5200 \\
\hline
HK & 1 yr & ~88 & ~3.7 & ~~1900 & ~~2800 \\
HK & 13.6 yr & ~6.7 & ~0.28 & ~~140 & ~~210 \\
DUNE & 1 yr & ~190 & ~9.0 & ~~150 & ~~1600  \\
DUNE & 13.6 yr & ~14  & ~0.69 & ~~11 & ~~120
\end{tabular}
\caption{\label{table:fluxexp} Required fluxes in unit of $10^{-7}$ cm$^{-2}$ s$^{-1}$ with which our reference points become sensitive in various experiments.
}
\end{table}

In both $e$-scattering and $p$-scattering cases, we expect to observe two or three separate signatures, which are not expected in usual neutrino scattering.
So it is fair to obtain the experimental sensitivity by requiring three signal events which correspond to the 95\% C.L. upper limit under the assumption of a null observation over a null background with Poisson statistics (see also Refs.~\cite{Dermisek:2016via,Dermisek:2014qca} for the related discussion), while we leave more systematic background analysis to the future work.
We list the minimum required fluxes of $\chi_1$ making our reference points sensitive in SK, HK, and DUNE in Table~\ref{table:fluxexp}.
Considering the fact that the typical flux of $\chi_1$ demanded in the minimal BDM setup is $\mathcal{O}(10^{-7})$ cm$^{-2}$ s$^{-1}$~\cite{Agashe:2014yua}, we see that $e$-ref. 2 is rather {\it promising}.
The other reference points can be also probed once we consider a modified BDM setup to increase the flux up to $\mathcal{O}(10^{-4})$ cm$^{-2}$ s$^{-1}$~\cite{Kong:2014mia,Alhazmi:2016qcs} or fixed target experiments with much higher intensity~\cite{future}.
Note that the sensitivities in HK 1-year is much better, compared to SK 13.6-year, mainly due to the bigger volume. For $p$-scattering we observe that the sensitivities increase in DUNE due to its remarkably lower $E_p^{\rm th}$.

\begin{figure}[t]
\includegraphics[width=0.494\linewidth]{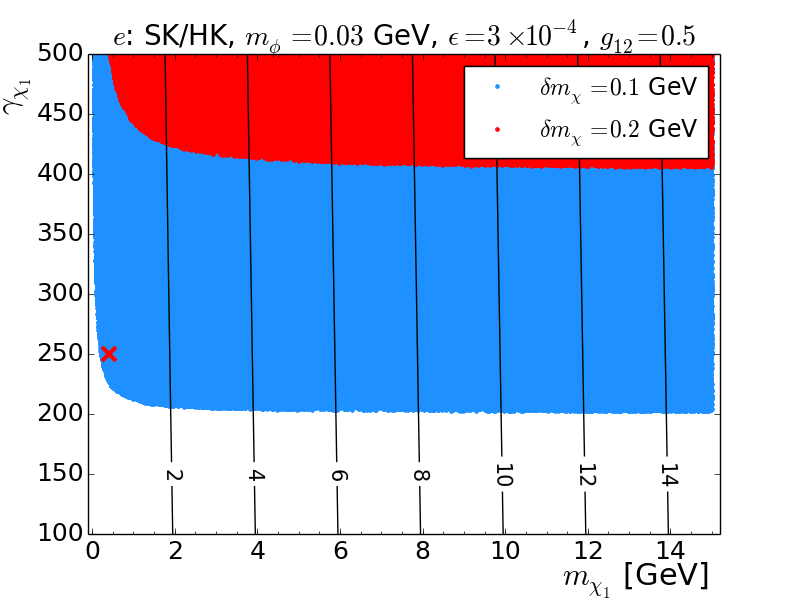}
\includegraphics[width=0.494\linewidth]{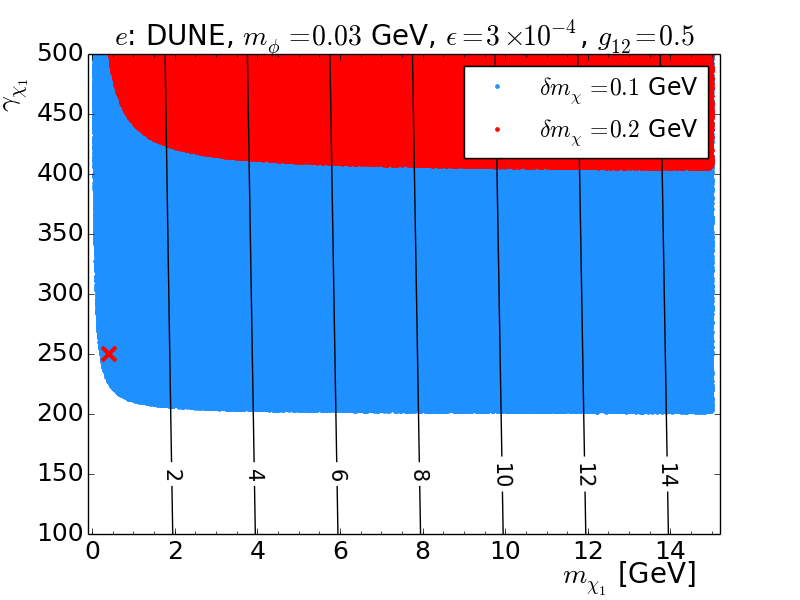}
\includegraphics[width=0.494\linewidth]{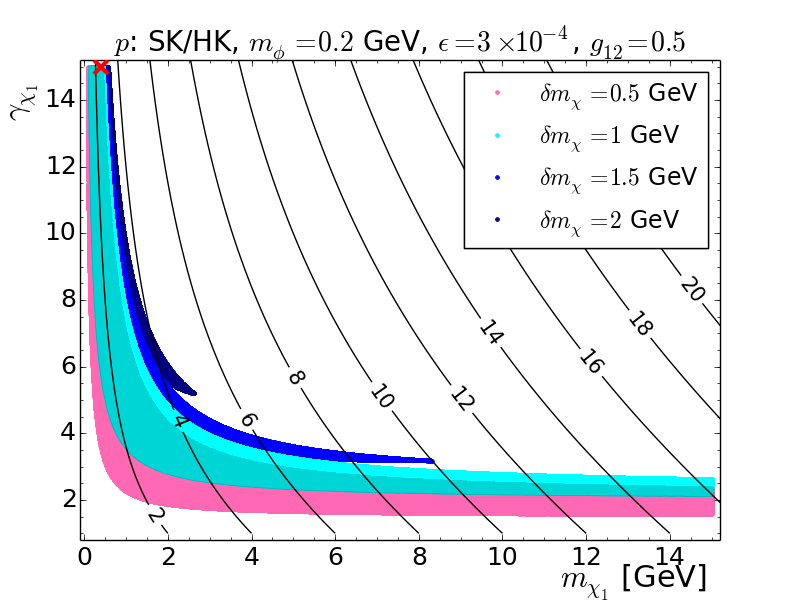}
\includegraphics[width=0.494\linewidth]{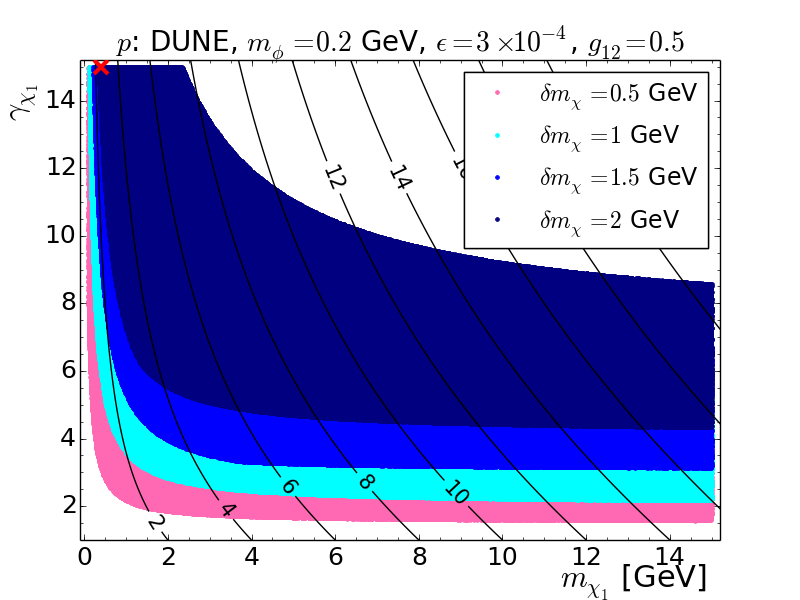}
\caption{\label{fig:scan} $e$-scattering (upper panels) and $p$-scattering (lower panels) parameters in SK and HK (left panels) and DUNE (right panels) for $m_\phi = 0.2$ GeV. For $e$-scattering we scan for $\delta m_\chi = 0.1, 0.2$ GeV while $\delta m_\chi = 0.5, 1, 1.5, 2$ GeV for $p$-scattering. Black contours show maximally allowed $m_{\chi_2}$ values in GeV and the red crosses show our reference parameters.
}
\end{figure}

We finally conduct a parameter scan to check the viability of our signal processes in a wider range of space. Fixing $m_{\phi} = 0.03\, (0.2)$ GeV, $\epsilon = 3 \times 10^{-4}$, and $g_{12} = 0.5$, we obtain the allowed parameter region \footnote{For $p$-scattering, we define the allowed parameter region satisfies $\sigma_{\rm R.O.I} / \sigma_{\rm tot} > 0.5$ where $\sigma_{\rm tot}$ is the total cross section in the whole kinematically allowed range of $|\vec p_p|$ and $\sigma_{\rm R.O.I}$ is the region of interest in which the momentum transfer to the target proton is not too large to break up the proton (practically, $|\vec p_p| \in [E_p^{\rm th}, 1.8\,{\rm GeV}]$).} of $e$($p$)-scattering in $m_{\chi_1}$ vs. $\gamma_{\chi_1}$ plane for $\delta m_{\chi} = 0.1$, 0.2 GeV (0.5, 1, 1.5, 2 GeV) and show them in the upper (lower) panels of Fig.~\ref{fig:scan}. The left (right) panels are for SK/HK (DUNE). The black contours represent the maximally accessible $m_{\chi_2}$ for a given set of $m_{\chi_1}$ and $\gamma_{\chi_2}$ (see Eq.~\eqref{eq:mchi2} as well). Minimally required $\chi_1$ fluxes for our signal to be sensitive in each experiment are an order of magnitude smaller than (of the same order as) those for $e$-scattering ($p$-scattering) in Table~\ref{table:fluxexp}. The red ``X'' points denote the reference points: $e$-ref.1 and $p$-ref.1.


\paragraph*{{\bf Conclusions and outlook.}}

In this letter, we proposed a novel DM detection strategy for the models with non-minimal dark sector involving a heavier unstable particle. Once boosted DM inelastically scatters off target material and produces a heavier dark sector particle, secondary signatures may arise in associated with the target recoil, as shown in Fig.~\ref{fig:scenario}.
This signal feature clearly differs from relativistic neutrino scattering events, which offers a {\it new paradigm} in probing non-minimality of the dark sector.
We also investigated the detection prospects of relevant DM signals at large volume neutrino detectors, and found promising. Similar analyses will be straightforwardly applicable to future fixed target experiments.

It is possible to study more complicated signatures such as multi-step cascade decays although we employed the simplest secondary process in this letter. Furthermore, we expect proactive utilization of the knowledge from collider phenomenology due to the similarity of the proposed DM scenario with typical collider signatures, when detectors are designed and implemented accordingly in the future.
As a concluding remark, we strongly encourage DM-related intensity-frontier collaborations (e.g., LBNF/DUNE, SHiP, and T2HKK)
to pay their attention to the proposal in this letter as possible physics to pursue.

\section*{Acknowledgments}
We thank K.~Agashe, M.~Aoki, J.-H.~Huh, D.~Kim, K.~Kong, K.~Matchev, T.~Saab, C.~S.~Shin, M.~Son, and J.~Yoo for insightful/useful discussions.
We appreciate Brookhaven Forum 2015 for its encouraging environment enabling us to initiate this project, and CETUP* (Center for Theoretical Underground Physics and Related Areas), Santa Fe LHC summer workshop 2016, and Focus Workshop on Particle Physics and Cosmology by IBS-CTPU for their hospitality during the completion of this work.
DK was supported in part by DOE Grant DE-SC0010296, and is presently supported by the Korean Research Foundation (KRF) through the CERN-Korea Fellowship program.
JCP is supported by the National Research Foundation of Korea (NRF-2016R1C1B2015225) and the POSCO Science Fellowship of POSCO TJ Park Foundation. 
SS was supported in part by DOE Grant DE-SC0010120.

\end{document}